\documentclass[a4paper,11pt,onecolumn]{article}
\usepackage{latexsym}
\usepackage{amssymb, amsmath}
\usepackage{graphicx}
\usepackage{dcolumn}
\usepackage{bm}
\def\be{\begin{equation}}
\def\ee{\end{equation}}
\def\beq{\begin{eqnarray}}
\def\eeq{\end{eqnarray}}
\def\bn{\begin{eqnarray*}}
\def\en{\end{eqnarray*}}

\def\P{\Phi}
\def\p{\phi}

\def\a{\alpha}
\def\b{\beta}
\def\s{\sigma}

\def\d{\delta}
\def\D{\Delta}
\def\g{\gamma}
\def\t{\theta}
\def\T{\Theta}

\def\si{\psi}

\def\n{\nu}
\def\m{\mu}

\def\t{\theta}

\input epsf

\begin{document}
\title{NATURALNESS AND ELECTRO-WEAK SYMMETRY BREAKING\footnote{Lectures at the
 {\it Advanced  School: From Strings to LHC-II}, Bangalore, India,
 December 11-18, 2007.}}
\author{
Romesh K. Kaul \\
The Institute of Mathematical Sciences, \\
Chennai 600 113, India.\\
kaul@imsc.res.in
}

\date{~}

\maketitle

\begin{abstract}
The Principle of Naturalness of  small parameters of a theory is reviewed.
While quantum field theories constructed from gauge fields and fermions
only are natural, those containing elementary scalar fields are not.
In particular the Higgs boson mass in  the Standard Model of electro-weak 
forces is not stable against radiative corrections. Two old canonical
solutions of this
problem are: (i) where  the Higgs boson is a fermion-antifermion composite 
(technicolour solution) or otherwise (ii) we need  supersymmetry
to protect the mass of  elementary Higgs boson from possible
large radiative corrections. In recent years some  other mechanisms 
for electroweak symmetry breaking have been under intense investigation.
These include the little Higgs
models and the gauge-Higgs unification models where the Higgs boson 
is the zero mode of  the extra-dimensional component of a higher dimensional
gauge field.  Naturalness issues of such models are also briefly
reviewed.

\end{abstract}

\section{Naturalness dogma}

When confronted with experiments, the Standard Model (SM) of
fundamental forces has proven to be very robust  
both in its general structure as well as in every detail tested so far. 
However, a satisfactory understanding of the origin of  electro-weak 
symmetry breaking mechanism has been an ever elusive problem. In 
the SM  electro-weak  symmetry $SU(2)_L\times U(1)_Y$ is spontaneously
broken to $U(1)_{EM}$ through a non-zero vacuum expectation value of an
$SU(2)$ doublet of elementary scalar fields. There are also 
other possibilities of achieving this goal. Besides the obvious 
requirement that the weak
gauge bosons $W^{\pm}_\m$ and $Z^0_\m$ acquire the requisite
masses, there is one guiding principle, known as the {\it
Naturalness Principle}, which is largely adopted in working out
any new mechanism for the symmetry breaking.

The dogma or principle of naturalness expresses the belief that 
a small parameter in Nature can not be an accident. It must 
be associated with a symmetry. This is in contrast to an 
anthropic principle.

The naturalness principle is best formulated through what 
can be called 't Hooft's doctrine of naturalness \cite{thooft}:

\vspace{0.2cm}

{\it At any energy scale $\m$, a set of parameters, $\a^{}_i(\m)$ describing 
a system can be small, if and only if, in the limit 
$\a^{}_i(\m) \rightarrow 0$ for each of these parameters, 
the system exhibits an enhanced symmetery.}

\vspace{0.2cm}

Weakly broken symmetry ensures that the smallness of a parameter 
is preserved against possible perturbative disturbances.

Let us analyse naturalness of various parameters in some of 
the quantum field theories that we come across in particle physics:

\vspace{0.3cm}

{\bf (1) Quantum electrodynamics is a perfectly natural theory.}
This theory describes electromagnetic interaction of charged 
fermions, electron $\lambda^{}_e$,  muon $\lambda^{}_{\m}$, etc:
\bn
{\cal L}^{}_{QED} ~= ~-~{\frac 1 4} ~F^{\m\n}_{} F^{}_{\m\n} ~+~
{\sum^{}_{f=e,\m, ...}} {\bar \lambda}^{}_f \left[ i\g^\m_{} \left(
\partial^{}_\m - ieq^{}_f A^{}_\m \right) -m^{}_f \right] \lambda^{}_f
\en
\noindent Here the electromagnetic coupling $e$, the electron mass 
$m^{}_e$, the muon mass $m^{}_\m$ etc. can all be independently 
small. The smallness of $m^{}_e$ (or $m^{}_\m$) is
protected by the fact that, in the limit $m^{}_e\rightarrow 0$
(or $m^{}_\m \rightarrow 0$), we have an additional symmetry, 
the chiral symmetry which corresponds to separate conservation of the 
left- and right-handed electron-like leptons. No surprise that 
all the perturbative corrections to electron mass due to quantum 
fluctuations are small. These are in fact just proportional to $m^{}_e$
itself; the self-energy diagrams have only logarithmic divergence. 
Also, $e \rightarrow 0$ enhances the symmetry; it implies no 
interaction; hence the particle number of each type is conserved 
in this limit. One-loop quantum correction to the electromagnetic 
coupling $e$ is indeed logarithmically divergent and is proportional 
to $e^2_{}$.

Same discussion is valid for electromagnetic interaction of all 
other charged fermions and their masses, in particular for quarks.

\vspace{0.3cm}

{\bf (2) Quantum Chromodynamics (QCD) is also a perfectly natural theory.}
It describes the colour dynamics of  $SU (3)$ triplet quarks
$\lambda_{}^i$ $ (i = 1, 2, 3)$ and an octect of gluons $A^a_\m$ 
$ (a = 1, 2, 3, ...8)$:
\bn
{\cal L}^{}_{QCD}  = ~-~ {\frac  1 4}~ F^{a\m\n}_{} F^{a}_{\m\n} ~ +~
{\bar \lambda}^i_{} \left[ i\g^\m_{} \left( \d^{}_{ij} \partial^{}_\m
- {\frac  {ig} 2} (T^a_{})^{}_{ij} A^a_\m \right)  - m\d^{}_{ij} \right]
\lambda^j_{}
\en
\noindent where $T^a _{}$ are the eight generators of $SU (3) $
algebra in triplet representation. Here again the colour coupling 
constant $g$ is natural, because in the limit $g \rightarrow 0$, 
there is enhanced symmetry. It reflects no interaction, hence 
particle number of each type in conserved in this limit. 
One-loop perturbative quantum corrections to the coupling 
constant $g$ have only logarithmic divergence and are 
proportional to $g^2_{}$. Also mass parameter $m$ is natural
because in the limit $m \rightarrow 0$ we have the chiral symmetry
which is preserved by pertubative quantum corrections;
such  corrections to $m$ are logarithmically divergent and
are proportional to $m$ itself.

\vspace{0.3cm}

{\bf (3) Quantum theories involving interacting elementary scalar 
fields are not natural.} This has to do with the fact that the mass 
of an elementary scalar field is not associated with any 
approximate symmetry. Consider a self-interacting theory of a 
real scalar field:
\bn
{\cal L}^{}_{scalar} =~ {\frac 1 2}~ \partial^\m_{} \p~ \partial^{}_\m \p
~-~ {\frac {m^2_{}} 2}~ \p^2_{} ~-~ {\frac {\lambda} {4!}}~ \p^4_{}
\en

At the classical level, the limit mass $m \rightarrow 0$ does lead to 
scale invariance; but at quantum level scale symmetry 
is broken. Thus smallness of the scalar mass can not be protected 
against perturbative quantum corrections. In fact such corrections 
appear with quadratic divergences. For example, at one loop level, 
such a correction comes from the diagram in Fig 1:
\vskip0.2cm
\centerline{\epsfbox{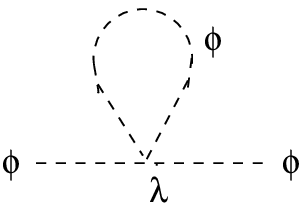}}
\centerline{ Fig.1. One-loop radiative correction to scalar mass}
\vskip0.2cm
\noindent and it is:
\bn
\d m^2_{} \sim \lambda\int^{\Lambda^2_{}} dk^2_{} {\frac {k^2_{}} {k^2_{} -m^2_{}}}
\sim \lambda~ \Lambda^2_{}
\en

On the other hand, the other parameter of this theory, 
namely the $\p^4_{}$ coupling $\lambda$ is natural. This is so because, 
in the limit $\lambda \rightarrow 0$, we have a free scalar theory, 
which indeed has higher symmetry.

\section{Naturalness of electro-weak model}

Next let us consider a more general theory containing 
an elementary charged scalar field where local $U (1)$ gauge invariance is 
spontaneously broken through a nonzero vacuum expectation value 
of the scalar field. The theory describes a complex scalar 
field $\p$  and a left-handed fermion  $\si_L$ interacting with a $U (1)$ 
gauge field $A^{}_\m$ and  a right-handed fermion $\si_R$ which is neutral
with respect to this gauge symmetry and also  Yukawa coupling of the fermions
with the scalar field:
\beq
{\cal L} =-{\frac 1 4}~F^{\m\n}_{}F^{}_{\m\n}+i{\bar \si}^{}_L\g^\m_{}
\left( \partial^{}_\m -ieA^{}_\m \right) \si^{}_L + i{\bar \si}^{}_R
\g^\m_{}\partial^{}_\m \si^{}_R \cr
+\left( \partial_{}^\m +ie A^\m_{} \right) \p^*_{} \left( \partial^{}_\m
-ieA^{}_\m \right) \p  \cr
+ \m^2_{} \p^*_{} \p -\lambda (\p^*_{} \p)^2_{} - Y\left[\p{\bar \si}^{}_L \si^{}_R
+\p^*_{} {\bar \si}^{}_R \si^{}_L \right] \label{higgsaction}
\eeq
\noindent This theory is invariant under the following gauge transformations:
\bn
A'_\m = A^{}_\m +{\frac 1 e}~ \partial^{}_\m \t, ~~~~ \p' = e^{i \t}_{} \p,
~~~~\si'_L = e^{i\t}_{} \si^{}_L, ~~~~\si'_R = \si^{}_R.
\en

The field theory described by the Lagrangian density (\ref{higgsaction})
has many of the features of the Standard Model of electro-weak
interactions. Besides a spontaneous breaking of symmetry and a
Higgs mechanism it also has a Yukawa coupling for fermion and
scalar fields which leads to the mass for  fermion like
in the SM. Unlike the SM this theory has
anomalies in the $U(1)$ gauge current. This can be cured by adding
another  left-handed fermion with opposite $U(1)$ charge
to the already included fermion. However the discussion of naturalness
issues below does not depend on this.

Of the four parameters, the dimensionless couplings  $e$ and 
$Y$  are independently  natural as in the limit $e \rightarrow 0$,  
$Y \rightarrow 0$ we have 
no gauge interaction and no Yukawa interaction respectively, 
and hence enhanced symmetries. Indeed perturbative quantum corrections 
to these parameters are proportional to themselves.
But in the case of dimesionless
parameter $\lambda$ situation is different. Even in the limit
$\lambda \rightarrow 0$ at tree level, presence of gauge and Yukawa 
interactions induce quantum corrections, say at one loop level, to generate
a non-zero $(\phi^* \phi)^2$ interaction in the effective potential
with a coefficient
$M e^4+ N Y^4$ ($M$ and $N$ are some constants) due to gauge 
and fermion fields in the loops.
This puts restrictions on how small the effective coupling
$\lambda$ can be. It can not be very much smaller than 
the gauge and Yukawa couplings. 

The dimensionful parameter $m$ deserves special attention.
Notice due to the wrong sign of $\p^*_{} \p$  term $(\m^2_{} > 0)$, 
the $U (1)$ symmetry is broken. The potential $V (\p, \p^*_{}) = 
-\m^2_{} \p^*_{}\p + \lambda (\p^*_{}\p)^2_{} $  has a ring of minima 
given by $\p^*_{} \p = v^2_{}/2 \equiv \m^2_{} /(2\lambda) $. Thus 
expanding the scalar field about its minimum value as 
$\p = (v+ H) e^{i\s}_{} / {\sqrt 2}$  and rotating the phase away 
by absorbing the field $\s$ (would be Nambu-Goldstone mode) into 
the massive vector field as its longitudinal component through 
a change of variables: $A_\m^{} \rightarrow A^{}_\m - \partial^{}_\m \s/e$
and also $\si^{}_L \rightarrow e^{-i\s}_{} \si^{}_L $, we obtain a 
theory of a massive vector field $A^{}_\m$, a massive scalar field $H$
(so called Higgs field) and a massive Dirac fermion  $\si$
with their masses given by:
\bn
m^{}_A = ev, ~~~~ m^{}_H = {\sqrt {2\lambda}} ~ v, ~~~~ m^{}_{\si} = 
{\frac 1 {\sqrt2}}~ Y v.
\en

Now in the limit ~$v \rightarrow 0$ ~(or equivalently ~$\m \rightarrow 0$), 
we do have enhancement ~of ~symmetry  classically;  in this ~ limit we have 
~(i) scale invariance, ~(ii) restored $U(1)$ gauge symmetry and also, 
(iii) since Dirac fermion becomes massless, chiral symmetry 
corresponding to the separate conservation of left- and right-handed 
fermions. Yet this does not make the vacuum expectation value $v$ or 
the mass parameter $\m$ of the original Lagrangian (\ref{higgsaction}) 
natural. None of these classical symmetries can protect the vector 
field mass $m^{}_A$, Higgs field mass $m^{}_H$ or fermion mass $m^{}_{\si}$. 
This is so because quantum fluctuations, through the well known 
Coleman-Weinberg mechanism, break all these symmetries by 
inducing a non-zero vacuum expectation value of the scalar field 
even if it were zero to start with in this limit of the classical 
theory. Thus at quantum level, there is no enhancement of symmetry 
in the limit where classical vacuum expectation value $v$ tends to zero. 
It is important to contrast this situation with the case of QED 
discussed above where electron mass $m^{}_e$ is protected because chiral 
symmetry in the limit $ m^{}_e \rightarrow 0$ is an exact 
quantum symmetry. 

Indeed in the present context, perturbative quantum 
corrections to each of the masses $m^{}_A$, $m^{}_H$ and $m^{}_{\si}$
have quadratic divergences. That is, corrections to the (mass)$^2$ of 
these fields are proportional to the square of cut-off $\Lambda$. 
The vacuum expectation value $v$ of the scalar field also receives 
radiative corrections which are quadratically divergent. These come from
radiative diagrams of the type in Fig 2.1.

\centerline{\epsfbox{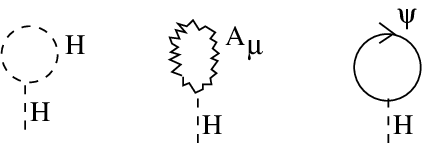}}
\centerline{Fig.2.1. One-loop radiative corrections to the vacuum 
expectation value}

\vspace{0.3cm}

\noindent This one-loop correction is given by: $v^2_{1-loop} = v^2_{}+$
${\frac 1 {16\pi^2_{}}}
(P \lambda $ $ +Qe^2_{} -R Y^2_{}) \Lambda^2$ where the three terms come from three
diagrams of Fig 2.1. Notice in the limit $v \rightarrow 0$, one-loop
vacuum expectation value is not zero. This is Coleman-Weinberg mechanism 
of radiative symmetry breaking.

Next
for the Higgs field $m^2_H$ receives a correction at one-loop level 
as: $\d m^2_H \sim \a~ \Lambda^2_{}$ with $\a ={\frac 1 {16\pi^2_{}}}
(A \lambda +Be^2_{} -C Y^2_{})$ where $A$, $B$, and $C$ are numerical 
constants and respectively correspond to the Higgs field, 
gauge field and fermion field going through the loop as in Fig. 2.2.
\vskip0.2cm
\centerline{\epsfbox{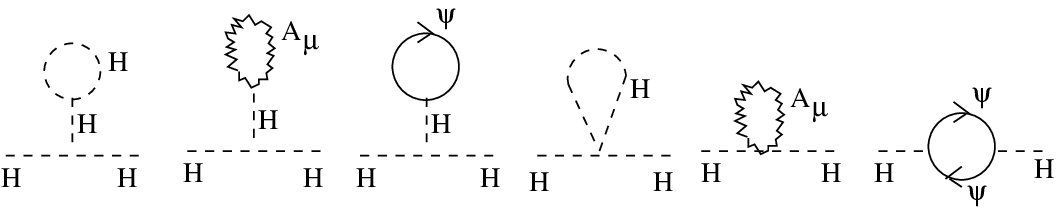}}
\centerline{Fig.2.2. One-loop radiative corrections to the Higgs mass}
\vskip0.2cm

Clearly this discussion holds for the Standard Model of electro-weak 
forces also. Here $SU(2)^{}_L \times U(1)^{}_Y$ symmetry is spontaneously 
broken to the electromagnetic symmetry $U(1)^{}_{EM}$ due to the 
vacuum expectation value of a doublet scalar field resulting in 
a massive physical scalar field, the Higgs field H. In the process the three 
weak gauge bosons $W^{\pm}_{}$ and $Z^0_{}$ also become massive and  
quarks and leptons acquire their masses through Yukawa couplings. 
Thus in this model also, radiative corrections to Higgs mass $m^{}_H$
diverge quadratically as the internal momentum in the loop becomes 
large. Then our computations break down for loop momenta $p^2_{} \sim \Lambda^2_{}$
where the cut-off $\Lambda$  is the energy scale up to which the 
SM is an adequate description of Nature. We may write, say 
the one-loop correction to Higgs boson mass due to quantum fluctuations 
of a size characterised by the scale $\Lambda$  as:
\be
\d m^2_H \sim \a~ \Lambda^2_{}
\ee

Similar corrections obtain for the vacuum expectation value of 
the scalar field as well as for the masses of vector bosons $W^{\pm }_{}$ 
and $Z^0_{}$ and also fermions. Thus, for $m^{}_H \sim 100~ GeV$, and 
coupling $\a \sim (100)^{-1}_{}$, the requirement that this mass 
does not receive large radiative corrections, $\d m^2_H \sim m^2_H$,
we have:
\bn
\Lambda^2_{} \sim {\frac {\d m^2_H} {\a}} = {\frac {(100~ GeV)^2_{}} {(100)^{-1}_{}}}
= (1000~ GeV)^2_{}
\en
\noindent That is, the cut-off $\Lambda \sim  1~ T eV$. Naturalness 
of electro-weak theory breaks down at this scale. This is not any problem 
if there is no physics beyond $1~ T eV$; that is, if there are no 
elementary particles heavier than this scale, or there is no physical
characteristic mass scale beyond this value. But in general, there is 
no reason to expect that this is so. For example, if the idea of 
grand unification for electro-weak and strong forces is valid, 
there is a physical scale, the grand-unification scale which 
is much larger than $1~ T eV$ . In particular, in the $SU (5)$ Grand 
Unified Theory (GUT) we have two widely separated scales: the GUT 
scale $M^{}_{GUT}$ $\sim 10^{16}_{}~ GeV$ where the GUT gauge group $SU(5)$
spontaneously breaks down to the SM group $SU(3)^{}_C \times SU(2)^{}_L
\times U(1)^{}_Y$ and the scale of electro-weak physics $M^{}_{EW} 
\sim  100~ GeV$ where the electro-weak gauge group $SU(2)^{}_L\times  U(1)^{}_Y$
breaks down to $U(1)^{}_{EM}$. These two levels of symmetry breaking 
are achieved through vacuum expectation values of two sets of 
elementary scalar fields. It was in this context first that it was 
realized that the radiative corrections do not allow the two scales 
to be maintained at such widely separated values \cite{gildner}. 

In fact, not only  quadratically divergent radiative diagrams, but also
some times certain kind of logarithmically divergent  diagrams contribute
large corrections to small masses. Such diagrams come with large 
coefficients proportional to the larger
mass scale of the theory. These kind of {\it large} logarithmic divergent
diagrams are typically present in GUTs. To understand the origin
of such radiative corrections, consider
a theory of two interacting scalar fields, $\P$ and $\p$, which
have two vastly separated vacuum expectation values generated
by some suitable potentials: $<\P> = F$ and $<\p> = f$ with
$F >> f$. Expanding about their expectation values $\P = F +H$
and $\p = f+h$, we have two massive scalars fields with their
masses $m_H \sim F$ and $m_h \sim f$. A possible interaction
of the type $ a ~\P^2 \p^2$ in the original potential would lead,
after shifting the fields by their vacuum expectation values, 
to effective three-point vertices of the form $ a~ F H hh$
with a  large dimensionful coupling $a~ F$. Such
vertices will in turn lead to a {\it large} logarithmically
divergent radiative correction to the mass
of the light scalar field $h$ as
\bn
\d m^2_h \sim a^2 ~F^2 ~\ln \Lambda^2
\en

\noindent from a radiative diagram shown in Fig.2.3:

\vskip0.1cm
\centerline{\epsfbox{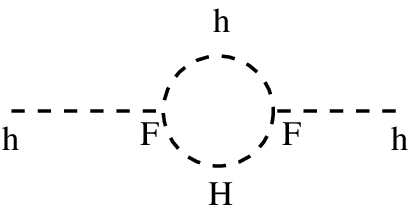}}
\centerline{Fig.2.3. {\it Large} logarithmically divergent 
radiative correction }
\vskip0.2cm

Such corrections along with those large corrections
from quadratically divergent diagrams would destabilise
the mass of lighter scalar field. This is a generic
feature of all such theories of scalar fields with two 
widely separated scales. In particular, it is true of GUTs:
perturbative quantum corrections tend to draw the smaller electro-weak
scale $M_W \sim 10^2~ GeV$ towards the GUT scale 
$M_{GUT} \sim 10^{16} ~GeV$. This problem known as the {\it gauge hierarchy
problem} of grand unified theories is due to the fact that field
theories containing elementary scalars are not natural.

Not only this, there is yet another much larger and important 
physical mass scale $M^{}_{Planck} = 10^{19}_{}~ GeV$ in Nature 
which is associated with quantum gravity. This would imply that 
the radiative corrections would draw the masses of electro-weak theory 
to this high scale and hence their natural values would be 
$\sim 10^{19}_{}~ GeV$ and not the physical values characterised by 
the SM scale of $100~ GeV$! What this quantum instability of
the Higgs potential strongly suggests  that there has 
to be some new physics beyond $1~ T eV$ such that the SM 
with its characteristic scale of $100 ~GeV$ stays natural beyond 
this scale. Or otherwise there can not be any fundamental scale 
in Nature beyond $1~ T eV$; in particular the scale of four 
dimensional gravity, $M^{}_{Planck}$ can not be a fundamental scale; 
but it should be derived from some new physics characterised by 
a fundamental scale of $1~ TeV$ only \footnote{Some of the higher
dimensional theories exhibit such a feature. It
may also be possible to achieve this, if starting from a classically
scale invariant
theory of all the  interactions including gravity, both
the weak scale and the Planck scale are generated by the quantum
effects which break this invariance weakly, ensuring that
any scalar fields which can acquire large vacuum expectation
values are only weakly coupled  to the SM sector\cite{foot}}.

\section{Composite Higgs boson as a solution of naturalness
problem}

Earliest discovered solution to the naturalness problem of electro-weak 
theory is one where the Higgs particle is not an elemetary particle 
but a spin-less composite of a fermion and anti-fermion. This is 
the Technicolour solution \cite{weinberg, techni}. The constituents 
of the Higgs boson are to be held together by a new strong force. 
Known interactions like colour $SU(3)^{}_C$ are not sufficiently strong 
at the electro-weak energy scale, a new much stronger interaction is 
needed to obtain the required bound states\footnote{While composites 
made up of the light quarks 
such as $u, ~d$ and $s$ are not adequate,  much heavier top quark
with its strong coupling to the electro-weak sector, may serve the purpose.
It is worthwhile to ponder over this possibility of 
symmetry breaking  where the Higgs boson
would be a $t$ ${\bar t}$ bound state. But unfortunately, this minimalist
idea of a composite Higgs boson  does 
not work because it requires
a top quark of mass much heavier that the experimentally observed value 
\cite{bardeen}.}. 
This interaction should 
exhibit confinement. For no better reason than the fact QCD is a 
confining gauge theory, the new interaction is postulated to be 
a QCD type theory, though, operative at a higher scale of about $1~ TeV$,
where its running coupling constant becomes of order unity. The Higgs 
particle would then be built in the same fashion as pions are in QCD. 
If we were to probe the Higgs particle with energies greater than 
$1 ~TeV$, we would see it not as an elementary scalar particle, 
but instead as a fermion and an anti-fermion. Except for a few, 
most of the physical states of this new theory would have high 
enough masses characterised by the high energy scale of this theory.

Historically a variety of names were proposed for these new 
fermions such as metafermions ('t Hooft), heavy fermions (Weinberg), 
hyperfermions (Eichten and Lane), technifermions (Susskind) and 
the new interaction experienced by them was called metacolour, 
heavy colour, hypercolour and technicolour respectively. The last 
nomenclature has survived over years.

Since theories with fermions and gauge fields like QED and QCD as 
indicated earlier, are natural, Technicolour theories do not 
suffer from Naturalness problem.

Besides technicolour forces, the techniquarks are also supposed 
to experience electro-weak interactions. Simplest technicolour model 
can be build in terms of an $SU(2)^{}_L$ doublet $(U, D)^{}_L$ of 
techniquarks and two right-handed singlets $U^{}_R$ and $D^{}_R$. Such 
theory exhibits an additional global flavour symmetry 
$SU(2)^{}_L \times SU(2)^{}_R \times  U(1)^{}_V$ which is broken 
to $SU(2)^{}_{L+R} $ $\times U(1)^{}_V$ through the condensation 
of the techniquarks leading to three Nambu -Goldstone bosons 
(technipions) $\pi^a_{TC}$. These then in turn induce required 
masses for the weak interaction bosons $W^{\pm}_{}$ and $Z^0_{}$. 
These masses fix the analog of the pion decay constant $F^{}_{TC\pi}$
of the new interaction, defined in terms of matrix elements of 
the associated spontaneously broken axial current $J^\m_{5a}$  by:
\bn
<0|J^\m_{5a}|\pi^{}_{TCb}(q)> ~=~ iq^\m_{} ~F^{}_{TC\pi}~ \d^{}_{ab}
\en
\noindent The $W$ boson mass is related in a model independent 
way to the technipion decay constant as:
\bn
M^{}_W ~=~ {\frac 1 2}~g^{}_2~F^{}_{TC\pi}
\en
\noindent where $g^{}_2$ is the $SU(2)^{}_L$ coupling constant. 
Because of the isospin symmetry of the strong interaction, 
the relation $M^{}_W = M^{}_Z \cos \t^{}_W$ holds as long as 
electric charge is conserved.

Now the technipion decay constant can be fixed by relating 
it to the Fermi constant $G^{}_F$ of weak interaction as follows:
\bn
F^{}_{TC\pi} ~=~ {\frac 2 {g^{}_2}}~ M^{}_W ~=~\left( {\sqrt 2}G^{}_F
\right)^{-{\frac 1 2}}_{} ~=~ 250~ GeV.
\en
\noindent Since techni-QCD is only a scaled up QCD, the technipion 
decay constant $F^{}_{TC\pi}$ and ordinary QCD pion decay constant 
$f^{}_{\pi}$ are related as $F^{}_{TC\pi}/f^{}_{\pi}$ $\sim  \Lambda^{}_{TC}/
\Lambda^{}_C$ where $\Lambda^{}_C$ and $\Lambda^{}_{TC}$ are the scales 
at which the QCD and techni-QCD coupling constants respectively 
become strong. Taking $\Lambda^{}_C/f^{}_{\pi}$ $\sim 2$, we have 
the techni-QCD scale $\Lambda^{}_{TC} \sim  0.5~ TeV$.

This does yield a possible dynamical explanation of the origin of 
masses of electro-weak bosons $W^{\pm}_{}$ and $Z^0_{}$, but does not 
provide for the masses of ordinary quarks and leptons which in 
the Standard Model are generated through Yukawa couplings of 
these fermions with the elementary scalar field. In the technicolour 
framework, to generate these masses a new gauge interaction, 
the Extended Technicolour (ETC) is introduced. The gauge bosons 
of this new interaction, with their masses in the range $10$ - $100~ TeV$, 
connect the ordinary quarks and leptons to techniquarks; thus 
providing a mechanism for masses for quarks and leptons.

There are some serious phenomenological difficulties with 
such scenarios. These include not enough suppression of flavour 
changing neutral current effects, heavier Higgs particle, 
presence of large anomalous contributions to the $Zb{\bar b}$ vertex 
and large contributions to the Peskin-Takeuchi $S$, $T$ and $U$ parameters. 
But these difficulties may be the problem of specific model for 
the dynamics of force responsible for holding the constituent 
fermions together in the composite Higgs boson. It may be that a 
QCD-type model for this force is not adequate.
In particular, the relation $F^{}_{TC\pi} /f^{}_{\pi} \sim 
\Lambda^{}_{TC} /\Lambda^{}_C$ implied by the scaled up  QCD-type model for the new
interactions  is too restrictive.

\section{Supersymmetric solution}

That supersymmetry also provides a solution to the naturalness 
problem of electro-weak theory was realized about 27 years 
ago \cite{kaul, witten}. This option retains the elementarity of 
the scalar field. While composite Higgs boson is a non-perturbative 
solution to the problem, supersymmetry provides a perturbative solution. 
An elementary property of quantum field theory which gives an extra 
minus sign for the radiative diagram with a fermionic as against 
a boson field going around in a loop allows for the possibility 
that naturalness violating effects due to bosonic and fermionic 
quantum fluctuations can be arranged to cancel against each 
other. For this to happen the various couplings of bosons and 
fermions have to be related to each other in a highly
restrictive manner. Further, for such a cancellation to hold at 
every order of perturbation theory, a symmetry between bosons 
and fermions would be imperative. This is what supersymmetry 
indeed does provide.

In most of the supersymmetric field theories, troublesome 
quadratic divergences and also the {\it large} logarithmic divergences
are independently absent \cite{kaul}. This happens due to exact 
cancellation of such divergences between graphs with bosonic 
and fermion fields going around in the loops. While supersymmetric 
theories with a non-Abelian symmetry are always free of such 
divergences, those with $U(1)$ symmetry do have radiative corrections 
with quadratic divergence which are proportional to the sum of $U(1)$
charges of all the fields. So theories which have $U(1)$ charges 
adding up to zero do not have quadratic divergences. 
Supersymmetrized Standard Model is one such theory. This would, 
in particular, be also the case for any theory with an $U(1)$ symmetry 
that can be embedded in a non-Abelian group.

Also in supersymmetric field theories where gauge symmetry is 
broken spontaneously through non-zero vacuum expectation value of 
a scalar field, like in the SM, the limit when this vacuum 
expectation value goes to zero does indeed lead to an enhancement 
of symmetry even at the quantum level, provided, if there is 
a $U(1)$ symmetry present in the theory, $U(1)$ charges add up 
to zero. Quantum corrections do not spoil this symmetry; 
unlike the non-supersymmetric case, Coleman-Weinberg mechanism 
does not produce radiative violation of the symmetry. The Higgs 
boson mass is natural here and also so are the masses 
generated for gauge bosons through Higgs mechanism and 
the fermion masses generated through Yukawa couplings to scalar 
field with non-zero vacuum expectation value.

Supersymmetry requires that bosons and fermions come in families 
\cite{susy}: photon has a fermionic partner, the photino; 
electron's bosonic partner is selectron; quarks have scalar 
partners squarks, etc. Similarly, if we are interested in gravity, 
spin $2$ graviton has a fermionic super-partner spin $3/2$ gravitino.

Exact supersymmetry would imply that all properties except 
the spin of particles in a supermultiplet are the same. Thus, the 
masses and couplings of super partners would exactly be same. 
This, however, is not seen to be the case in Nature, otherwise we 
would have, for example, already observed the super partner of electron, 
selectron with same charge and mass as the electron. Supersymmetry has 
to be broken in Nature. But this breaking should be such that the 
basic reason of naturalness does not get out of hand again. While 
particle and sparticle masses, $M^{}_{part}$ and $M^{}_{spart}$,
have to be different, with $M^{}_{spart}$ sufficiently high to have 
escaped detection till now, the cancellation of bosonic and 
fermionic radiative corrections need only be
up to the naturalness breakdown scale of the SM:
\bn
|\int^{\Lambda^2_{}} dk^2_{} {\frac {k^2_{}} {k^2_{} - M^2_{spart}}}
- \int^{\Lambda^2_{}} dk^2_{} {\frac {k^2_{}} {k^2_{} -M^2_{part}}}~| \cr
~\sim ~|M^2_{spart} - M^2_{part} |~ \ln \Lambda^2_{} ~\le ~
\left( 1~ TeV \right)^2_{}.
\en
\noindent So the supersymmetry breaking has to be such that 
quadratically divergent parts of the radiative corrections cancel, 
but logarithmically divergent contributions need not cancel 
exactly. Such situations are obtained if supersymmetry is broken 
spontaneously or by what are called soft-terms in the action.

\section{Naturalness of little Higgs models}

In recent times, an alternative symmetry breaking mechanism 
for electroweak theory has been proposed where the Higgs boson, 
though an elementary scalar particle, is a pseudo-Nambu-Goldstone boson 
\cite{littlehiggs}. Such a Higgs boson is massless at the tree 
level. The symmetry is explicitly broken by weakly coupled operators 
in the theory so that the Higgs boson acquires mass without 
generating any quadratically divergent contributions at one-loop 
level; it has only a logarithmically divergent correction at this 
level. It is hoped that this mass is protected by the global symmetry 
with which the Higgs boson is associated as a Nambu-Goldstone boson 
when this symmetry is spontaneously broken. 
The electro-weak $SU(2)^{}_L \times U(1)^{}_Y$ is embedded in a 
larger gauge group, simplest being $SU(3) \times U(1)$. 
Models based on such an idea are called Little Higgs Models
\footnote{The contents
of this Section have been developed with Gautam Bhattacharyya}.

To understand the underlying structure let us consider an 
$SU(3)$ gauge theory with two scalar fields $\Phi^{}_1$ and $\Phi^{}_2$,
each transforming as a complex triplet of the gauge group 
described by the Lagrangian density:

\be
{\cal L} = |\left( \partial^{}_\m+igA^{}_\m \right)\P^{}_1 |^2_{}
+|\left( \partial^{}_\m + ig A^{}_\m \right)\P^{}_2 |^2_{} - {\frac  1 2}~
tr ~F^{}_{\m\n} F^{\m\n}_{} - V(\P^{}_1, \P^{}_2) \label{LHaction}
\ee
\noindent where the potential is:
\be
V(\P^{}_1, \P^{}_2) ~=~ {\frac {\lambda^2_{}} 2} \left( \P^{\dagger}_1 
\P^{}_1 -f^2_{} \right)^2_{} +~ {\frac {\lambda^2_{}} 2} \left( \P^{\dagger}_2
\P^{}_2 - f^2_{} \right)^2_{} \label{potential}
\ee

This potential has two global $SU(3)$ symmetries acting on the two 
triplets which are broken to two $SU(2)$ symmetries through the 
tree level vacuum expectation values $< \Phi^{\dagger}_1\Phi^{}_1 >$
$=~< \Phi^{\dagger}_2\Phi^{}_2 > = f^2_{}$. This produces ten 
Nambu-Goldstone bosons, five each from the two spontaneous 
breakings $SU(3) \rightarrow SU(2)$. Further, the gauge couplings 
in the Lagrangian density above represent a weakly gauged 
vector $SU(3)$ such that the global $[SU (3)]^2_{}$
is explicitly broken to a diagonal $SU (3)$. Five of the 
Nambu-Goldstone bosons become the longitudinal components of 
the five gauge fields through the Higgs mechanism, 
leaving behind other five Nambu-Goldstone bosons which 
are massless at the tree level.

Now let us understand the naturalness properties of this 
model: is the Higgs particle mass so generated natural? 
In the original Lagrangian density (\ref{LHaction}), we have 
one dimensionful parameter, namely $f$ which is the tree level 
expectation value of the scalar fields $\Phi^{}_1$  and $\Phi^{}_2$. 
In the limit $f \rightarrow0$, classically we do have enhanced 
symmetry. But at quantum level, like any theory of elementary 
scalar fields, this is not so due to the Coleman-Weinberg mechanism. 
Hence the vacuum expectation values of the scalar fields, 
$< \Phi^{}_1 > $ and $< \Phi^{}_2 >$, are not protected. These 
do receive large, quadratically divergent, corrections due to 
quantum fluctuations at one loop itself.

For definiteness, we write the one-loop correction to the 
effective potential as:
\bn
\D{\cal L} = \lambda^2_{}\Lambda^2_{} \a(g^2_{}, \lambda^2_{}) \left(
\P^{\dagger}_1 \P^{}_1 + \P^{\dagger}_2 \P^{}_2 \right)
+~\b(g^2) |\P^{\dagger}_1 \P^{}_2|^2_{} \ln (\Lambda^2_{}/ F^2_{})
\en
\noindent where $\a(g^2_{},\lambda^2_{}) = {\frac 1 {16 \pi^2_{}}}
(Ag^2_{} + B\lambda^2_{})$ and $\b(g^2_{})={\frac 1 {16\pi^2_{}}}Cg^4_{}$
with $A$, $B$ and $C$ as numerical factors. The first term with 
quadratic divergence comes from diagrams (a), (b), (c) and (d) 
and the last logarithmically divergent term comes from the diagram (e) 
of Fig.3 below:
\vskip0.2cm
\centerline{\epsfbox{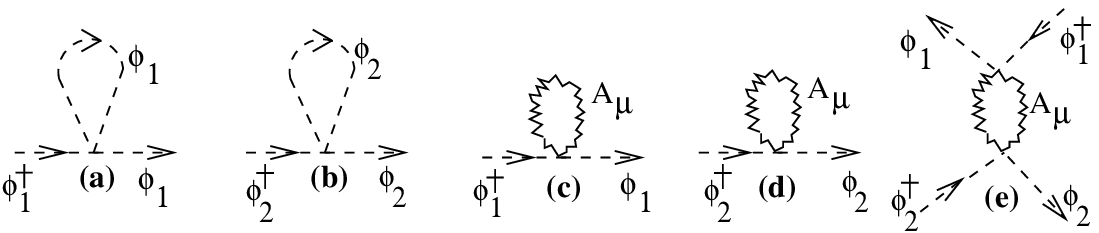}}
\centerline{Fig.3. One-loop radiative corrections}
\vskip0.2cm
Adding this to the tree level potential, we have one-loop 
effective potential as:
\be
V^{}_{1loop} = {\frac {\lambda^2_{}} 2} \left( \P^{\dagger}_1
\P^{}_1 -F^2_{} \right)^2_{} +~ {\frac {\lambda^2_{}} 2} \left( \P^{\dagger}_2
\P^{}_2 - F^2_{} \right)^2_{} - ~\b(g^2) |\P^{\dagger}_1 \P^{}_2|^2_{} \ln (\Lambda^2_{}/ F^2_{}) \label{1loopeff}
\ee
\noindent where $F$ is the one-loop corrected vacuum expectation 
value of the scalar fields:
\be
<\P^{\dagger}_1 \P^{}_1 >^{}_{1loop} = < \P^{\dagger}_2 \P^{}_2>^{}_{1loop}
= F^2_{} \equiv f^2_{} +\a(g^2_{}, \lambda^2_{}) ~\Lambda^2_{} \label{1loopvev}
\ee
\noindent Clearly the vacuum expectation values of the scalar 
fields are not protected; as expected their tree-level value $f$ does 
receive quadratically divergent corrections.

Now the triplet scalar fields can be expanded about their 
vacuum expectation value as:
\bn
\P^{}_1 = e^{i\T^a_{}T^a_{}/F}_{} \left( \begin{array}{c} 0 \\
0\\  {F+\eta^{}_1} \end{array}
\right), ~~~~~~~~\P^{}_2 = e^{-i\T^a_{}T^a_{}/F}_{} \left( 
\begin{array}{c} 0 \\
0\\  {F+\eta^{}_2} \end{array}
\right)
\en
\noindent where $T^a_{}$ are the $SU(3)$ generators and displaying 
only the left over five Nambu-Goldstone bosons in the exponent we 
write:
\be
\T^a_{}T^a_{} = {\frac 1 {\sqrt 2}} \left( \begin{array}{ccc}
0& 0& h^{}_1 \\
0& 0& h^{}_2 \\
h^*_1 & h^*_2 & 0 \end{array} \right) + {\frac \eta 4} \left( \begin{array}{ccc}
1 & 0& 0 \\
0 & 1& 0\\
0& 0& -2 \end{array} \right) \label{para}
\ee

The complex fields $h \equiv (h^{}_1, h_2^{})^T_{}$ are to be identified 
with the Standard Model doublet of scalar fields and $\eta$ is a 
singlet neutral real scalar field. At tree level the vacuum expectation 
value of the scalar fields $\Phi^{}_1$ and $\Phi^{}_2$ leaves an 
$SU(2)$ gauge symmetry unbroken; fields $h^{}_1$ and $h^{}_2$ and 
also $\eta$ have zero vacuum expectation values. All the five 
Nambu-Goldstone bosons $h^{}_1$, $h^{}_2$ and $\eta$ are massless 
at this level. But quantum corrections involving the gauge 
interactions of (\ref{LHaction}) represented by the last 
logarithmically divergent term in the one-loop effective potential 
(\ref{1loopeff}) above and also other such contributions 
from possible interactions with fermions of the theory which 
we have not displayed here, allows the doublet $h$ to pick up a 
non-zero vacuum expectation value and hence break the $SU(2)$ gauge 
symmetry in the standard way leading to three real fields becoming 
the longitudinal components of the $SU(2)$ gauge bosons leaving 
one massive real Higgs particle $H$. This follows immediately 
if we use the parametrisation (\ref{para}) to write the 
expansion:
\bn
|\P^{\dagger}_1 \P^{}_2|^2_{}=~F^2_{}\left[ -a h^{\dagger}_{}h + 
{\frac b {F^2_{}}} (h^{\dagger}_{}h)^2_{} + ..... \right] 
\en
\noindent where $a$ and $b$ are positive numerical coefficients. 
The first two terms of  effective potential (\ref{1loopeff}) do 
not contribute to the potential for $h$. Only contribution 
to this effective potential comes 
from the last term with logarithmic divergence so that: 
\bn
V^{}_{1loop} (h^{\dagger}_{}h) \sim -\b F^2_{}\left[ -a h^{\dagger}_{}h + {\frac b {F^2_{}}} (h^{\dagger}_{}h)^2_{} + ..... \right] \ln {\frac {\Lambda^2_{}}
{F^2_{}}}
\en

In order to have the SM symmetry breaking $\b$ has to be negative. 
This can be ensured by including fermions (for example a top quark)
with large Yukawa 
couplings $Y$ to scalars $\Phi^{}_1$ and $\Phi^{}_2$ which will 
also then, through graphs with fermion loops, contribute to 
both the quadratically and logarithmically divergent terms 
in (\ref{1loopeff}) and (\ref{1loopvev}) above so that 
$\a = {\frac 1 {16\pi^2_{}}}(Ag^2 + B\lambda^2 - EY^2)$ and 
$\b = {\frac 1 {16 \pi^2_{}}}(Cg^4$ $ - DY^4)$ in such a way 
that $\b  < 0$. Though we have not displayed them explicitly 
here, these fermions are essential to trigger the electro-weak 
symmetry breaking and hence have to be included in a 
complete description of the theory. Negative 
$\b$ allows the effective potential $V_{1loop} (h^{\dagger}h)$ to have 
a minimum at a non-zero vacuum expectation value $< h^{\dagger}h > $
$= [a/(2b)]F^2 $ $\equiv v^2_H /2$. Writing $h^{\dagger}h \sim {\frac 1 2}
(v^{}_H + H)^2 + ... $, we find that SM symmetry breaking takes 
place leading to a mass for the standard Higgs particle H as:
\be
M^2_H ~\sim~2(-a\b) ~F^2 \ln {\frac {\Lambda^2} {F^2}}
\ee
\noindent where $\b = {\frac 1 {16 \pi^2}} (Cg^4 -DY^4) <0$.

Though (mass)$^2$ of Higgs boson has only a logarithmic divergence 
at one loop level, it is also proportional to $F^2$ which does 
contain a quadratic divergence, $F^2 = f^2 +\a \Lambda^2$. This brings 
in the quadratic divergences into the Higgs mass.  Notice that this 
quadratic divergence  comes with a factor of the coupling constant $\a$. 
This is yet another example of a {\it large} logarithmically divergent
radiative correction in addition to the one that was mentioned earlier
in the context of the GUTs with widely separated scales in Section 2.

Now in order the mass of Higgs particle be around $100 ~GeV$, and for
$(-a\b) \sim (100)^{-1}$ and $ \a \sim (100)^{-1}$, we have 
$f^2 \sim  F^2 \sim (1~ TeV)^2$ and the cut-off $\Lambda  = 10~ TeV$. 

There is a phenomenological difficulty with regard to the weak gauge
boson mass in the little Higgs model we have outlined above. The quartic
coupling of the Higgs field $(h^\dagger h)^2$ is too small and hence
the weak gauge boson mass turns out to be too large, order $g_{EW}^{~} F$.
However, there are other little Higgs models, like the `Littlest Higgs model'
\cite{littlehiggs}  based on an
$SU(5)/SO(5)$ sigma model, which do not suffer from such a limitation.

Like the SM, little Higgs models also suffer from naturalness
problem. However, the scale of naturalness breakdown is about $10~ TeV$, 
an order of magnitude higher than that for the SM where, as discussed earlier 
in Section 2, it is only $1~ TeV$. This would mean that any new physics
mass scale beyond $10~TeV$, say  that associated with the
grand unification of strong and electro-weak forces,
would  destabilise this $ 10~ TeV$ scale thus giving
the electro-weak Higgs boson a huge radiative correction characterised by
the GUT scale. There is nothing  to protect the mass of this Higgs boson 
from receiving such a large radiative correction.
This is to be contrasted with the composite
Higgs boson and supersymmetric cases  discussed in Sections 3 and 4.
For example, despite the large value of the GUT scale, in
a supersymmetric framework, it does not destabilise the supersymmetry
breaking scale at $ 1 ~TeV$; the electro-weak Higgs boson does not receive
huge radiative corrections from  the GUT scale.  

Finally, in the little Higgs models also, to have 
naturalness beyond this $10~TeV$ scale, some other mechanism
needs to be invoked at this scale. This could be yet another
little Higgs mechanism operative at about $10~ TeV$ to push the 
naturalness scale up by an order of magnitude to $ 100~TeV$.
There would have to be a ladder of such successive mechanisms. 
Otherwise, 
supersymmetry or fermion-antifermion compositness of the Higgs 
particle  may appear at this scale. In particular, 
instead of $1~ TeV$ supersymmetry for the SM, Nature need have 
supersymmetry only at a higher scale of $10~ TeV$. Alternatively,
even for Higgs boson mass around $100~ GeV$, the compositness
scale for the forces binding new fermions into an effective
Higgs boson can be at a higher scale of about $10 ~TeV$ in contrast 
to the $1 ~TeV$ scale of  conventional technicolour models 
discussed in Section 3. This allows for the possibilities
of resolving the phenomenological difficulties faced by the 
old technicolour composite Higgs models.

\section{Higher dimensional theories }

There is yet another framework that addresses the electro-weak symmetry
breaking with  elementary scalar fields. In this framework,
the four dimensional scalar fields are identified with the zero
modes of extra-dimensional components of a higher dimensional
gauge potential \cite{fairlie, hosotani, others}. 
When the extra dimensional space is not simply connected (for example
if it is $S^1$), there are Wilson line phases $\t_H^{}$ associated
with the extra dimensional components of the gauge field, analog of
Aharanov-Bohm phase in quantum mechanics. The four dimensional fluctuations
of these phases are identified with Higgs scalar fields.  These are
massless at tree level and acquire masses through radiative
corrections as a finite-volume effect.  When we probe 
the Higgs boson with energies
of the order of $1~ TeV$, the extra dimensions open up and
we start seeing it as an  extra-dimensional component of a
higher dimensional gauge field.
This framework is known as {\it gauge-Higgs unification}.

We could start with a five dimensional gauge
theory containing fermions on a manifold $M^4 \times S^1$ with 
size of the compactified fifth dimension as $2\pi R$. 
Since the extra-dimensional component $A^a_5$ of the gauge fields are in 
the adjoint representation of the gauge group, the starting gauge group
in the higher dimensional theory has to be  large enough, 
say an $SU(3)$ or $SO(5)$ or $G_2$, to accommodate the four dimensional
Higgs field of the $SU(2)_L  \times U(1)$ SM model. The fifth components
of momentum $p^5$ of both the gauge fields and fermions are discrete.
These fields are expanded in terms of their Kaluza-Klein (KK) modes.
At low energies we have an effective four dimensional theory obtained by
integrating the action over the fifth dimension. This is a theory
of massless zero modes
of the fields and a tower of massive KK excitations for each field
with their masses characterised by the KK mass scale $m^{}_{KK} \sim 1/R$
and discrete values of the fifth component of  momentum.

Quantum effects induce an effective four dimensional 
potential for the adjoint representation
extra-dimensional components  of the gauge field thereby breaking
the gauge symmetry.
These calculations have to include the radiative effects from
all the various KK excitations. 
The answer would have divergences, naively even quadratic divergences. 
But these require special care with regularization. We have to 
adopt a regularization which allows for the
fact that the extra-dimensional components are the fifth
components of five dimensional
gauge fields and in the limit $R \rightarrow \infty$, where we have five
dimensional gauge symmetry, $A^{a}_5$ should be massless. This is
not any different from the fact the photon self energy graphs in
four dimensional electrodynamics have to be regularised in
a gauge invariant manner so that the photon  does not acquire
a quadratically divergent mass and thereby break the gauge symmetry;
only a wave function renormalization is allowed. Now in the present
context, if we adopt a reasonable regularization, the effective potential
for the extra-dimensional component of the gauge field
turn out to be such that
the bosonic (gauge) field loops tend to lead to a minimum of this
effective potential at  $<A^{a}_5> =0$ 
whereas the fermionic
loops tend to draw this vacuum value  away from zero. So if we
arrange sufficiently many fermions in the theory we can have a
radiatively generated symmetry breaking potential
with its minimum at a non-zero value
of order $ 1/R$. This breaks the  gauge symmetry.
This way of  breaking symmetry due to radiative corrections
in a compactified higher dimensional gauge theory is known as {\it Hosotani
mechanism}. The mass of Higgs boson so generated by the radiative
corrections is of order $1/R$. Further it appears that the Higgs boson 
mass may be finite to all orders in five dimensions suggesting 
that it is independent of the physics of the cut-off scale.

It is important to note that in this gauge-Higgs unification framework,
in the limit $R \rightarrow \infty$, the Higgs boson  mass goes to zero and
we have the five dimensional gauge symmetry.
It is this five dimensional gauge symmetry which protects the
smallness of Higgs boson mass.

However there are two difficulties with this framework:
(i) The first problem faced is that the fermions in higher dimensional 
theory lead to vector theories in the reduced effective 
four dimensional theory at low energies instead of a theory with 
chiral fermions. This can be cured making the extra dimension
have a non-trivial topology or allow for non-vanishing flux in 
the extra dimensions. We do have chiral fermions if the extra 
dimensional space is an orbifold, say $S^1/Z_2$, instead of a circle
$S^1$. The left-right asymmetry is achieved by appropriate orbifold 
boundary conditions so that the matter content of the SM is obtained
in the effective low energy theory.
(ii) Second problem is faced when contact is made with
the required masses of the electro-weak theory.
The mass scales  of the theory, namely
gauge boson masses $m^{}_W$ and fermion masses as well as KK 
modes mass scale $m^{}_{KK}$ 
are all related and are order $ 1/R$.
The Higgs boson mass is down by a factor of the weak coupling
constant, $m_H \sim {\sqrt {g^2_4/(4 \pi)}}~ m^{}_W$ where
$g_4$ is the effective four dimensional gauge coupling which
is related to the five dimensional gauge coupling $g_5$ as $g_4 =
g_5/{\sqrt {2 \pi R}}$. This makes the KK energy scale 
rather low so that the masses of the low lying KK modes are 
same as that of weak gauge boson.
In addition the mass of Higgs boson is  too low. This phenomenological
difficulty arises mainly  because the framework has been set up in flat
five dimensional spacetime and could be circumvented if the theory
was instead set in curved spacetime.

Both these problems get resolved by setting up the theory 
in the Randall-Sundrum (RS) warped five dimensional spacetime \cite{RS}.
In this space time extra dimensional space has the topology of
orbifold $S^1/Z_2$ of radius $R$. 
We have here a five dimensional anti-de Sitter space where the
fifth dimension is an interval $| y| \le \pi R$
and with $k^{-1}$ as the AdS curvature.
The boundaries
of the interval are at the  fixed points $y =0$ and $ y= \pi R$ where
two three-branes,  the so called Planck or UV brane and  TeV or IR brane 
respectively, are located. 
Induced metric on these boundaries differ by the exponential 
warp factor $ e^{\pi kR}$  generating widely separated effective scales.
At the Planck brane 
$ y=0$, the effective four dimensional mass scale is of order the Planck
scale $ k \sim M_{Pl}$. On the other hand, the effective mass scale at the 
other brane at $y= \pi R$ is $M_{Pl} ~e^{-\pi kR} \sim 1~ TeV$.
The low energy four dimensional effective action for the zero modes
of the extra-dimensional gauge fields  is such that that these
zero-modes are localised near the TeV-brane.
The large warp factor $ e^{\pi kR}_{}$
relates the Planck scale $M^{}_{Pl}$ to electro-weak scale $m^{}_W$
through $M^{}_{Pl}/ m^{}_W ~\sim e^{\pi k R}~\sim 10^{17}_{}$ for
$kR \sim 12$ where $k \sim M_{Pl}$ and $-k^2$ is the cosmological 
constant in the bulk five dimensional space time. 
Thus RS spacetime provides a natural bridge between the Planck scale 
and weak scale through the warp factor.  The radius of
compactification $ R$ here, unlike the case of flat spacetime
where it is  of order $(1~TeV)^{-1}$, is not large; it is
instead as small as $\sim M_{Pl}^{-1}$.

This model has been studied for various five dimensional gauge
groups like $SU(3)$ and $SO(5) \times U(1)_{B-L}$ \cite{agashe}
with the latter having  more satisfactory phenomenological
properties, particularly  those related to the physics of
neutral currents of the electro-weak theory.
The KK mass scale is given by $m^{}_{KK} = \pi k e^{-\pi kR}$
which, for $kR\sim 12$ and $k \sim M^{}_{Pl}$,  is of order $1.5~ TeV$.
The KK spectrum is not equally spaced unlike in the flat spacetime
case.  The gauge boson and Higgs boson masses are predicted to 
be \cite{agashe}: $m^{}_W \sim 100~GeV$ and $ m^{}_H \sim  
120-290 ~GeV$. This makes the KK excitations to be sufficiently
heavy and at the same time the gauge and Higgs boson masses
of acceptable values.

In the RS framework there is an alternative proposal where Higgs boson
is not the zero mode of a higher
dimensional component of a gauge field but instead is an elementary scalar
field. There are two versions of such theories: (i)
the SM fields including the Higgs field live on the TeV brane,
only gravity propagates in the five dimensional bulk \cite{RS};
and (ii) all the SM fields live in the bulk \cite{chang}.

In the former case, since effective momentum cut off
near this brane is warped down from the five dimensional cut off
$\Lambda_5 \sim M_{Pl} \sim 10^{19}~ GeV$  to  $ \Lambda_{eff} \sim
\Lambda_5~ e^{-\pi k R} \sim 1~TeV$, radiative corrections to the Higgs
boson mass are cut off by this value. However care needs to be
taken when all the contributions of KK modes of the five
dimensional graviton are added. The zero mode of the graviton is
the standard four dimensional graviton  which has the standard
gravitational coupling to the matter fields on the TeV brane
and this is  small as $1/M^{}_{Pl} \sim (10^{19}~ GeV)^{-1}$.
However the couplings of the  KK gravitons with their masses given by the
characteristic KK scale $M_{KK} \sim k ~e^{-\pi k R}$, are enhanced
by the wrap factor to ${\frac 1 {M^{}_{Pl}}}~ e^{\pi kR}$ which is
only $\sim (1~TeV)^{-1}$. It is this large coupling that allows for the
possibility that such KK gravitons of $1 ~TeV$ mass may be observed
at a TeV collider, LHC. But these large couplings also create
a possible problem for the radiative corrections to the Higgs boson
mass from the tower of KK gravitons. Such corrections to the Higgs boson
mass at the infra red brane are
\bn
{\frac {\d m^2_H} {\Lambda^2_{eff}}} = \sum^{}_{KK} \left( {\frac {e^{\pi kR}}
{M^{}_{Pl}}}\right)^2 \left( M^{}_{Pl} ~e^{-\pi k R} \right)^2
= \sum^{}_{KK} ( TeV)^{-2}~ (TeV)^2 \sim \sum^{}_{KK} ~ 1
\en

\noindent Indeed the effective four dimensional momentum cut off is
$\Lambda^2_{eff} \sim (1~TeV)^2$. Each term in this correction
is small, but there is a sum over all the KK modes. This sum has to be
done in a meaningful way so that result does not become large again.

In the second case, where the SM fields including the Higgs field
live in the bulk, the tree-level bulk mass of the scalar field has to 
be small so that gauge bosons get reasonable masses of order $100~ GeV$.
But radiatively its natural value is $\sim k$. This brings
in the gauge hierarchy problem back in to the game \cite{chang}. 
However introduction of supersymmetry again  would protect 
mass of  the Higgs  mode far from the TeV-brane \cite{pomarol}. 
Such a model at low
energy is more like the Minimal Supersymmetric Standard Model (MSSM).

This is indeed a very bold proposal which needs some special care.
One problem that requires to be addressed is the mechanism for
fixing the size of the extra dimensions. Particularly, this has to be
done taking in to account the gravitational sector by including
the five dimensional curved spacetime metric. The zero modes of this metric
give four dimensional metric $g_{\m\n} (x)$ and the radion field
$R(x)$ whose vacuum value is the size $R$ of extra dimension.
A possible proposal for the stablization of the size of  extra dimension
developed by Goldberger and Wise \cite{goldberger}
involves a massive five dimensional scalar field. Equation of motion
of this field is solved with appropriate boundary condition in the RS
background. The solution in then put back into the action and extra
dimension integrated out to get an effective potential for the
modulus $R$. The minimum of the potential fixes the stable size of the
modulus to a value $k <R> \sim 12$. However the back reaction of
the scalar field
need not be small and can spoil this stablization. Also scalar fields
mass being prone to  problems from large quantum corrections, this
needs some extra care.

An additional problem is that the cosmological constant continues to be
not a natural parameter and has to be fine tuned to its small value.

\vspace{0.4cm}

{\bf Acknowledgements:} We thank the organisers of the {\it Advanced School:
From Strings to LHC-II} for excellent hospitality at the Fire-flies Ashram,
Bangalore.

\end{document}